\pdfoutput=1
\documentclass[a4paper]{article}
\usepackage{authblk}
\usepackage{lmodern}

\title{The Origin of Motif Families in Food Webs}
\author[1]{Janis Klaise\thanks{Corresponding author: J.Klaise@warwick.ac.uk}}
\author[1,2]{Samuel Johnson}
\affil[1]{Centre for Complexity Science, University of Warwick, Coventry CV4 7AL, United Kingdom}
\affil[2]{Warwick Mathematics Institute, University of Warwick, Coventry CV4 7AL, United Kingdom}
\date{}

\usepackage{amsmath}
\usepackage{amsfonts}

\usepackage{graphicx}
\graphicspath{{img/}} 

\usepackage[utf8]{inputenc}

\usepackage{todonotes}

\usepackage{csquotes}

\newcommand*\mean[1]{\bar{#1}}

\newcommand*\enmean[1]{\left\langle{#1}\right\rangle}

\usepackage{longtable}

\usepackage{standalone}

\usepackage{bibunits}
\usepackage{hyperref}
\usepackage{doi}
\hypersetup{%
  hyperfootnotes=false,
  colorlinks=true,
  citecolor=blue,%
  urlcolor=blue,
  linkcolor=blue
}

\begin{document}
\maketitle

\begin{abstract}
Food webs have been found to exhibit remarkable \enquote{motif profiles}, patterns in the relative prevalences of all possible three-species sub-graphs, and this has been related to ecosystem properties such as stability and robustness. Analysing $46$ food webs of various kinds, we find that most food webs fall into one of two distinct motif families. The separation between the families is well predicted by a global measure of hierarchical order in directed networks -- trophic coherence. We find that trophic coherence is also a good predictor for the extent of omnivory, defined as the tendency of species to feed on multiple trophic levels. We compare our results to a network assembly model that admits tunable trophic coherence via a single free parameter. The model is able to generate food webs in either of the two families by varying this parameter, and correctly classifies almost all the food webs in our database. This establishes a link between global order and local preying patterns in food webs.
\end{abstract}

{\bf Keywords:} ecology, food web structure, network motifs, trophic coherence

\section{Introduction}
Food webs are abstract representations of which species consume which others in an ecosystem \cite{Paine1966,Pimm1982,cohen2012community}. In a network-based description, species are represented by nodes and their trophic interactions are represented by directed links, pointing from prey to predator \cite{Pimm1982,dunne2004network,drossel}. Much work has been devoted to understanding the origin and meaning of the particular trophic interaction patterns observed in these food webs \cite{may1973stability,pimm_food_1991,garlaschelli2003universal}. Faced with the complexity of whole food webs, many researchers have focused on the interactions among subsets of species, through the analysis of small, connected subgraphs, or {\it motifs} \cite{camacho_quantitative_2007,paulau_motif_2015,borrelli_selection_2015,stouffer_evidence_2007,bascompte2005}.

The study of local interaction patterns via small network subgraphs \cite{itzkovitz_subgraphs_2003} first emerged in the study of neuronal and metabolic networks \cite{milo_network_2002,milo_superfamilies_2004}. The methodology of analyzing the relative prevalence of small subgraphs with respect to a well-posed null model for network assembly remains the main way to gain an understanding of the local structural properties of networks, including food webs \cite{borrelli_selection_2015,stouffer_evidence_2007}.

\begin{figure}[htbp]
  \centering
  \includegraphics[]{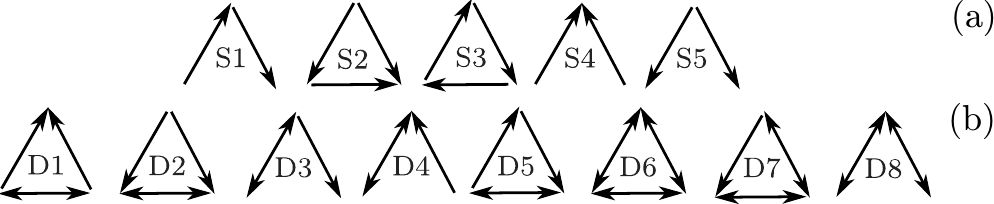}
  \caption{The 13 unique connected triads. These can be separated into two groups: (a) five triads, S1-S5, that only contain single links, (b) eight triads, D1-D8, that have double links (corresponding to mutual predation).}
  \label{fig:triads}
\end{figure}

In this study we focus on the $3$-node connected triads of which there are $13$ distinct ones (Figure \ref{fig:triads}). Many of these triads admit a straightforward interpretation in the context of food-webs \cite{stouffer_evidence_2007}. The eight triads D1--D8 have double links which correspond to mutual predation between two species. The five single link triads S1--S5 consist of some of the more basic building blocks of food webs. The triad S1 is a simple food chain \cite{cohen2012community,bascompte2005}, S2 represents omnivory (a predator preying on two species at different trophic levels) \cite{bascompte2005,polis1991}, triad S3 is a cycle (a relatively rare feature) \cite{stouffer_evidence_2007,polis1991}, and triads S4--S5 represent a predator preying on two species (apparent competition) and two predators sharing a prey species (direct competition), respectively \cite{borrelli_selection_2015}.

There are several competing hypotheses for the relative prevalence of these subgraphs in food webs. The prevailing hypotheses are that subgraphs emerge as a result of constraints (e.g the \enquote{niche dimension}) in the assembly of food networks \cite{camacho_quantitative_2007,stouffer_evidence_2007}, that functional importance leads to the observed structural patterns \cite{prill2005}, or that certain stability properties favour some subgraphs over others \cite{borrelli_selection_2015}.

Nevertheless, local structural patterns in complex networks are intimately related to global network properties \cite{vazquez_topological_2004,dominguez-garcia_inherent_2014}. A network metric called ``trophic coherence'' was recently introduced in order to capture the degree to which the nodes fall neatly into distinct levels \cite{johnson_trophic_2014,johnson_spectra_2015,dominguez-garcia_intervality_2016,klaise_neurons_2016}. In the context of food webs, these are the trophic levels, and high coherence corresponds to the species at one level consuming almost
exclusively species at the level immediately below (i.e. low omnivory). Trophic coherence was shown to be a major predictor of the linear stability of ecosystem models, as well as of a number of structural properties of empirical food webs \cite{johnson_trophic_2014}. It has also been related to the numbers of cycles in directed networks, and to the distribution of eigenvalues of associated matrices \cite{johnson_spectra_2015}.

Trophic coherence is a global structural property of directed networks, but it also places constraints on local topological features and on the prevalence of small subgraphs in particular. The extent to which these two properties, apparent at different scales, co-occur has
not been studied. In this paper we present evidence that the relative prevalence of three-vertex subgraphs in food webs can be explained by the level of trophic coherence in both empirical and model food webs. This result provides another viewpoint in the debate about the
origin of subgraph prevalences in food webs and further evidence of the importance of global organization in food webs \cite{johnson_trophic_2014}.

\section{Methods}
\subsection{Quantifying triad significance}
For any given network the exact number $N_k$ of any of the $k=1,\dots, 13$ connected three-vertex subgraphs (triads, Figure \ref{fig:triads}) is influenced by the network size and the degree distribution of the vertices. To test the statistical significance of any given triad $k$, the empirically observed number $N_k$ is compared against appearances of the same triad in a randomized ensemble of networks serving as a null model \cite{milo_superfamilies_2004}. This comparison gives a statistical significance or $z$-score

\begin{equation}
  z_k=\frac{N_k-\enmean{N_k}_{\text{rand}}}{\sigma_\text{rand}},
\end{equation}
where $\enmean{N_k}_\text{rand}$ and $\sigma_\text{rand}$ are the randomized ensemble average and standard deviation for triad $k$, respectively. The $z$-score of triad $k$ thus measures the deviation of prevalence in the observed network with respect to the null model.

The $z$-scores of all $13$ triads can be summarized in a triad significance profile (TSP) which is a vector $\mathbf{z}=\{z_k\}$ with components $z_k$ for each triad $k$. Additionally, the normalized version of the TSP  is often used to compare networks of different sizes and link densities \cite{milo_superfamilies_2004}. This is given by

\begin{equation}
  \mathbf{\hat{z}} = \left\{\frac{z_k}{\sqrt{\sum_{k=1}^{13}{z_k^2}}}\right\}.
\end{equation}
%

The randomization procedure used to obtain the randomized ensemble statistics is a matter of choice. A careful selection of null model is important to discern between real effects and artefacts present in the TSP \cite{beber_artefacts_2012}. In our analysis, we follow the configuration model (CM) prescription \cite{Newman2001,newman2010networks}, and preserve the number of incoming and outgoing links for each node (the degree sequence) while randomizing links via a Markov chain Monte Carlo switching algorithm \cite{milo_network_2002,milo_superfamilies_2004}. This preserves both the total number of nodes (species) and the links (trophic interactions) in the network. The generation of randomized networks and counts of triads was carried out with \textit{mfinder}, the algorithm used by Milo {\it et al.} in their seminal work on network motifs \cite{milo_network_2002,kashtan2002mfinder}.

It is important to emphasize that the TSP is a relative measure of which triads are over- and under-represented with respect to the null model provided by the randomized CM networks. The over-(under-)representation as indicated by a positive (negative) $z$-score indicates that these triads appear more (less) frequently than in the randomized networks but do not imply an absolute saturation (absence) of said triads. Nevertheless, the TSP is an adequate tool for comparing networks of different sizes and degree distributions.

\subsection{Comparing networks based on triad significance}
To quantitatively compare networks based on their triad significance profile, we use Pearson's correlation coefficient $r$ between the normalized $z$-score vectors $\mathbf{\hat{z}}^a$ and $\mathbf{\hat{z}}^b$ of networks $a$ and $b$, respectively \cite{stouffer_evidence_2007,milo_superfamilies_2004}. This is defined as

\begin{equation}
  r = \frac{\sum_{k=1}^{n} \left(\hat{z}_k^a-\mean{z}^a\right)\left(\hat{z}_k^b-\mean{z}^b\right)}{\sigma_{\mathbf{\hat{z}}^a}\sigma_{\mathbf{\hat{z}}^a}},
\end{equation}
where
\begin{equation}
  \mean{z}^a = \frac{\sum_{k=1}^{n}\hat{z}^a_k}{n}
\end{equation}
and
\begin{equation}
  \sigma_{\mathbf{\hat{z}}^a} = \sqrt{\sum_{k=1}^{n}\left(\hat{z}_k^a-\mean{z}^a\right)^2}
\end{equation}
are the mean and the standard deviation of the normalized $z$-score vectors; $a$ and $b$ specify the networks, $k$ is an index over the triads and $n=13$ is the total number of triads.

With this definition a value of $r$ close to $1$ indicates that the two networks have very similar TSPs and thus patterns of over- and under-represented triads, a value close to $0$ indicates no similarity, and a value close to $-1$ indicates anti-similarity -- i.e. triads over-represented in one network will typically be under-represented in the other (and vice versa).

Comparing the empirical networks between themselves is straightforward as we just calculate the $r$-coefficient pairwise for the $z$-score vectors of all $46$ food webs in our database. On the other hand, for comparison with the model (described in Section \ref{sec:model}), for each empirical network we fit the model to the data and generate $1000$ instances of a model network and then compute the $r$-coefficient of the empirical $z$-score vector and the average $z$-score vector of the model-generated ensemble.

\subsection{Trophic coherence}
Trophic coherence is a global metric for directed networks that characterizes how \enquote{layered} the network is \cite{johnson_trophic_2014,johnson_spectra_2015}. It measures the extent to which we can separate nodes into distinct groups so that any given group receives incoming links from just one other group and has outgoing links to another, different group of nodes. In the context of food webs, it measures the overall tendency of species to feed on multiple distinct trophic levels.

For each species $j$ in the network, we define its \textit{trophic level} $s_j$ as the average trophic level of its prey, plus one \cite{johnson_trophic_2014,LEVINE1980195},

\begin{equation}\label{eq:s}
  s_j = 1+\frac{1}{k_j^\text{in}}\sum_i a_{ij}s_i,
\end{equation}
where $k_j^\text{in}=\sum_i a_{ij}$ is the number of prey of species $j$ (also known as the \textit{in-degree}) and $a_{ij}$ are entries of the adjacency matrix $A$ of the food web. Here the convention is that the directed trophic links point from prey $i$ to predator $j$.

Because of the recursive nature of Equation \ref{eq:s}, to assign a trophic level to every node in the network two conditions must hold. First, there must be at least one node of zero in-degree -- we call such nodes \textit{basal}; and second, every node in the network must be reachable by a path from at least one basal node. Food webs satisfy both conditions,\footnote{In any ecosystem there are always basal species without prey (autotrophs), and every species is part of a food chain that starts with a basal species.} so the linear system defined by
Equation \ref{eq:s} has a unique solution. Without loss of generality we assign $s_j=1$ for all basal species, as is the convention in ecology.

\begin{figure}[htbp]
  \centering
  \centerline{\includegraphics[width=\linewidth]{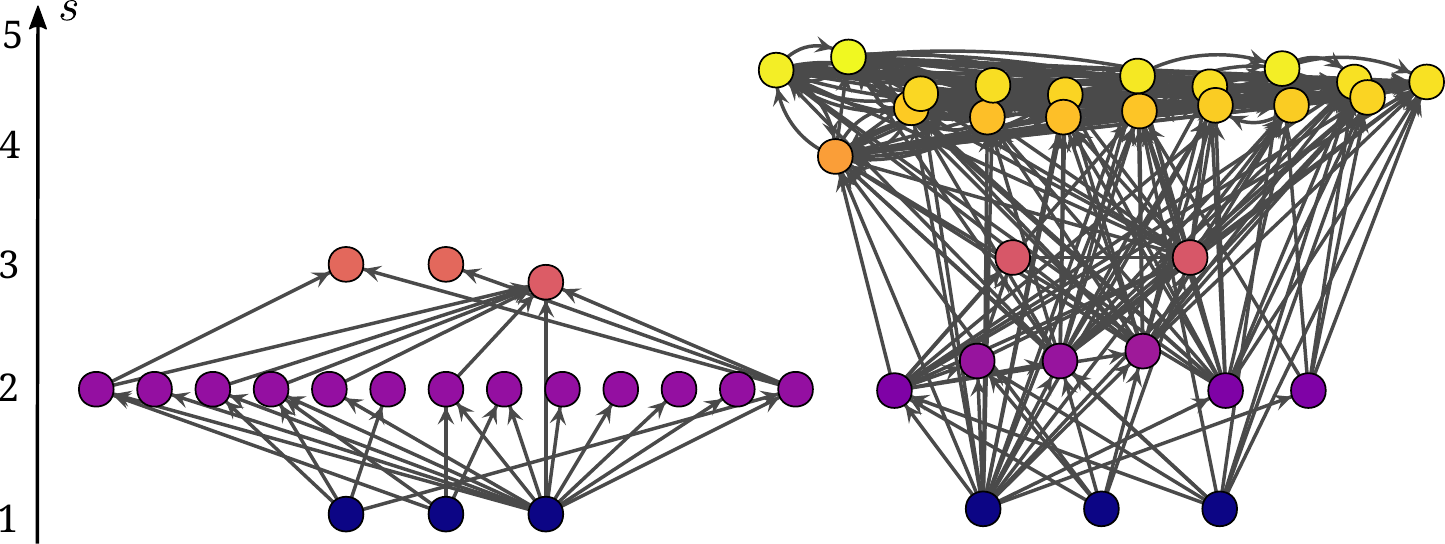}}
  \caption{Examples of different degrees of trophic coherence in food webs. Left: Crystal Lake (Delta) -- a highly coherent network with $q=0.17$, note that only one node prevents the network from being perfectly coherent ($q=0$). Right: Coachella Valley -- an incoherent network with $q=1.21$, note the high number of nodes falling between integer trophic levels due to the complex patterns of trophic links.}
  \label{fig:example}
\end{figure}

We define the \textit{trophic distance} associated to link $a_{ij}$ in the network as the difference between the trophic levels at the endpoints, $x_{ij}=s_j-s_i$. Note that this is not a distance in the mathematical sense as it can take negative values. Denote by $p(x)$ the distribution of trophic distances as measured on a network. This will have mean $\enmean{x}=1$ by definition and a standard deviation $q=\sqrt{\enmean{x^2}-1}$ which we will call the \textit{trophic incoherence} parameter.

The trophic incoherence parameter is thus a measure of the homogeneity of the distribution $p(x)$. For perfectly coherent networks we have $q=0$, which translates to having only integer valued trophic levels and all species feeding on prey only one trophic level below their own. In this case the network is perfectly structured, or layered, as there are distinct groups of \enquote{herbivores} feeding only on basal species, \enquote{predators} feeding only on \enquote{herbivores} and so on. For less coherent networks, $q>0$ indicates a less ordered trophic structure, where trophic levels take fractional values and species tend to prey on a broader
range of trophic levels. See Figure \ref{fig:example} for examples of coherent and incoherent food webs.

\subsection{Model with tunable trophic coherence}\label{sec:model}
Various mathematical models of food webs have been proposed to capture and explain different aspects of food webs \cite{cohen1985,williams2000simple,cattin2004phylogenetic,stouffer2005,stouffer2006}, but the main models still fail to capture the full complexity of empirically observed structures \cite{johnson_trophic_2014,allesina2008,williams2008}. We introduce a model for food webs that allows us to adjust the incoherence parameter $q$ by means of fitting a single free parameter. The model is a generalization of the Preferential Preying Model (PPM) introduced in Ref. \cite{johnson_trophic_2014}, with the improvement that it can generate bidirectional links and cycles of higher order, thus producing more realistic networks. In the following we denote by $B,N$ and $L$ the number of basal nodes, total nodes and links in the network respectively, all parameters to be fitted using the
empirical network data.

We begin with $B$ basal nodes and no links. We assign trophic levels $s=1$ to all basal nodes. We then add $N-B$ new nodes to the network sequentially according to the following rule. For each new node $j$, pick exactly one prey $i$ at random from among all the existing nodes in the network, thus creating a link from $i$ to $j$. In doing so, we define the temporary trophic level of node $j$ as $\hat{s}_j=1+\hat{s}_i$. After this procedure finishes, we have a network of $N$ nodes and $N-B$ links, and each node has a (temporary) trophic level $\hat{s}_i$.

Once all $N$ nodes are created, we add the remaining links to the network to bring the expected number of links up to $L$. The links are chosen among all possible pairs of nodes $\left(i,j\right)$ (we exclude connections to basal nodes), with a probability $P_{ij}$ that decays with the (temporary) trophic distance $\hat{x}_{ij} = \hat{s}_j-\hat{s}_i$ between them. Specifically, we set

\begin{equation}\label{eq:prob}
  P_{ij}\propto\exp{\left(-\frac{\left(\hat{x}_{ij}-1\right)^2}{2T^2}\right)},
\end{equation}
where $T$ is a free parameter which sets the degree of prey diversity between multiple trophic levels. This form of probability ensures that the most likely links to be created are between adjacent (temporary) trophic levels. The probabilities in Equation \ref{eq:prob} are normalized so that the expected number of links in the final network is $L$.\footnote{In principle, the link addition process can be amended so that the final number of links is exactly $L$
not just on average. However, this can be detrimental because in some situations, e.g. for low $T$ and high link density, the exact number of links $L$ may not be attainable without distorting the probability distribution set by Equation \ref{eq:prob}.}

At the end of the network creation procedure the trophic levels need to be recalculated according to Equation \ref{eq:s} as the addition of new links will have changed the network topology, and the trophic levels in the final network need not correspond to the temporary integer valued trophic levels.

\begin{figure}[htbp]
  \centering
  \includegraphics[]{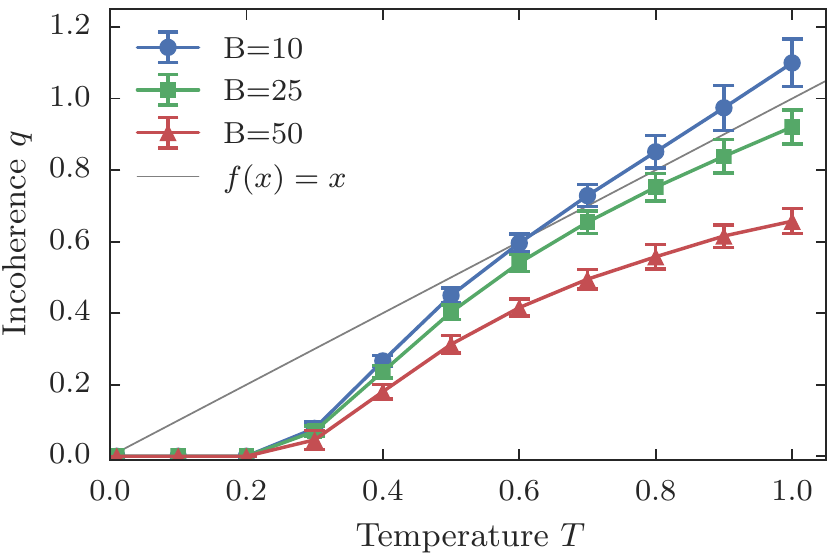}
  \caption{Dependence of the incoherence parameter $q$ on the temperature parameter $T$ in an ensemble of networks generated by the model. The networks in the ensembles have $N=100$ nodes $B$ of which are basal and average non-basal degree $\enmean{k}=L/(N-B)=10$. The averages are computed over at least $1000$ networks and error bars are one standard deviation of the sample.}
  \label{fig:qvT}
\end{figure}

The free parameter $T$ is analogous to temperature in statistical physics and sets the amount of deviation from a perfectly coherent network. For $T=0$, only links between adjacent (temporary) trophic levels are allowed which results in the incoherence parameter $q=0$. In this case the temporary trophic levels coincide with the actual trophic levels as the addition of links does not change the initially assigned trophic levels. As $T$ is increased, links between a wider range of (temporary) trophic levels become more probable, so we expect $q>0$ and increasingly more random networks. A sample dependence of $q$ on $T$ is shown in Figure \ref{fig:qvT}. The model exhibits a monotonic dependence of the incoherence parameter $q$ on temperature $T$ which provides a basis for fitting the model to empirical food webs given the empirically observed $q$. We also find that the level of incoherence that is achieved at any given temperature depends on $B/N$, the ratio of basal species to all species. We will further explore this relationship in Section \ref{sec:omnivory}.

To fit the model to the food web data, we provide as input the number of basal species $B$, the number of total species $N$, and the number of links or trophic interactions $L$. We then use stochastic root finding to find the value of the temperature parameter $T$ that results in an ensemble of networks whose incoherence parameter $q$ is centred about the empirical incoherence parameter as measured from the food web topology.

\section{Results}
\subsection{Motifs in empirical food webs}\label{sec:motifs}
\begin{figure}[htbp]
  \centering
  \includegraphics[]{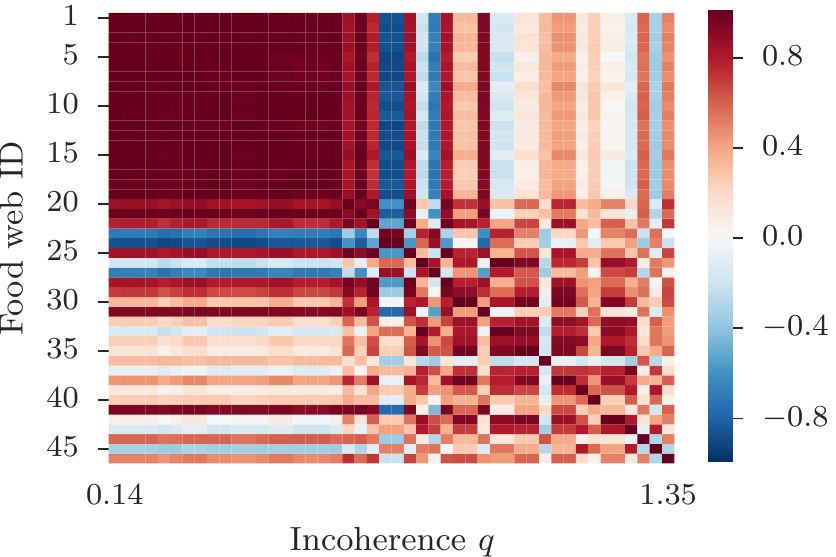}
  \caption{Pearson's correlation coefficient of the triad significance profiles (TSP). The coefficient is measured pairwise between all pairs of empirical food webs. Warmer colours indicate greater similarity while colder colours indicate dissimilarity. The food webs are arranged according to increasing incoherence parameter (left to right and top to bottom).}
  \label{fig:heatmap}
\end{figure}

\begin{figure}[htbp]
  \centering
  \includegraphics[]{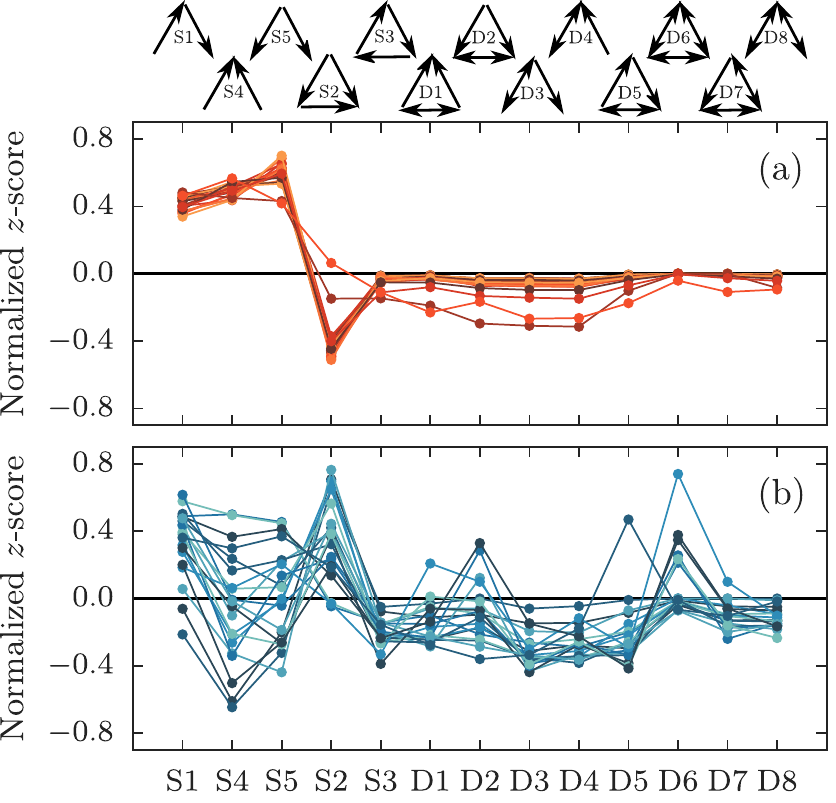}
  \caption{Triad significance profiles (TSP) as measured by the normalized $z$-score of the two groups of foodwebs. (a) Food webs in the first family (ID 1--22) with low incoherence parameter $q$ characterized by an over-representation of triads S1, S4 and S5 and an under-representation of triad S2. (b) Food webs in the second family (ID 23--46, except 30, 35, 40 and 43) with high incoherence parameter characterized by an over-representation of triad S2.}
  \label{fig:tsp}
\end{figure}

We study the triad significance profile (TSP) in $46$ empirical food webs from a variety of environments: marine, freshwater (river and lake) and terrestrial (See supplementary material for details). The results are summarized in Figures \ref{fig:heatmap} and \ref{fig:tsp}.

Figure \ref{fig:heatmap} shows the pairwise Pearson correlation coefficients of the triad significance profiles between all $46$ food webs. The food webs are arranged by increasing incoherence parameter $q$ so that more coherent food webs are assigned a lower ID. Red hue or warmer colours indicate a larger coefficient, while blue hue or colder colours indicate an anti-correlation in the TSPs.

We see that roughly two families of food webs emerge with similar TSPs. The first family (roughly ID 1-22) is characterized by relatively high coherence (low incoherence parameter $q$), for which the similarities in the TSPs are very high ($r\geq0.8$).

There is a second family of food webs, characterized by a high incoherence parameter $q$, that also show high similarities in their TSPs. Membership to this second family is not as clear as there is a tighter core of food webs belonging to it, with a periphery that only shares some similarities. There are also a number of sporadic food webs that do not appear to fall into either of the two families.

The change in family membership as the incoherence parameter $q$ increases indicates a qualitative change in the respective TSPs of the food webs. To better investigate what might be responsible for this change, we look closer at the bulk behaviour of the TSPs for the two families. Figure \ref{fig:tsp} shows the normalized profiles of Family 1 (top plot, ID 1-22) and Family 2 (bottom plot, ID 23-46, excluding  the sporadic food webs ID 30, 35, 40, 43).

We first consider Family 1. The bulk behaviour of food webs in this family is characterized by an over-representation of triads S1, S4 and S5, as well as an under-representation of triad S2. We should find the pattern of under-representation of triad S2 (which represents omnivory) unsurprising, since food webs belonging to this family have a low incoherence parameter $q$, which limits the ability of species to feed on multiple different trophic levels. Equally, the over-representation of triads S1, S4 and S5 is to be expected as these are the only three triads out of 13 that can arise in a hypothetical food web with $q=0$, which is a value close to the empirical values of $q$ for food webs in this family. The double link triads D1-D8 are all under-represented or close to even, in agreement with our expectations.

Family 2 is more interesting. Here the triads S1, S4 and S5 no longer follow a strong pattern of over-representation and the double link triads D1-D8 are not always under-represented. The most distinguishing feature, however, is the bulk over-representation of triad S2, in stark contrast to Family 1. We will argue in Section \ref{sec:omnivory} that this is the main feature that separates the two food web families.

This pattern of food webs belonging to two distinct families based on the under- or over-representation of triad S2 was already alluded to in \cite{stouffer_evidence_2007}, however it was in disagreement with the predictions of the generalized cascade \cite{stouffer2006} and niche \cite{williams2000simple} models which can only produce food webs where S2 is over-represented \cite{stouffer_evidence_2007}. In Section \ref{sec:omnivory} we present results from our model introduced in Section \ref{sec:model} which show that it is possible to change the pattern of under-representation to over-representation of triad S2 by increasing the incoherence parameter $q$, thus providing evidence that trophic coherence can naturally give rise to two food web families characterized by low or high prevalence of omnivory, respectively.

\begin{figure}[htbp]
  \centering
  \includegraphics[]{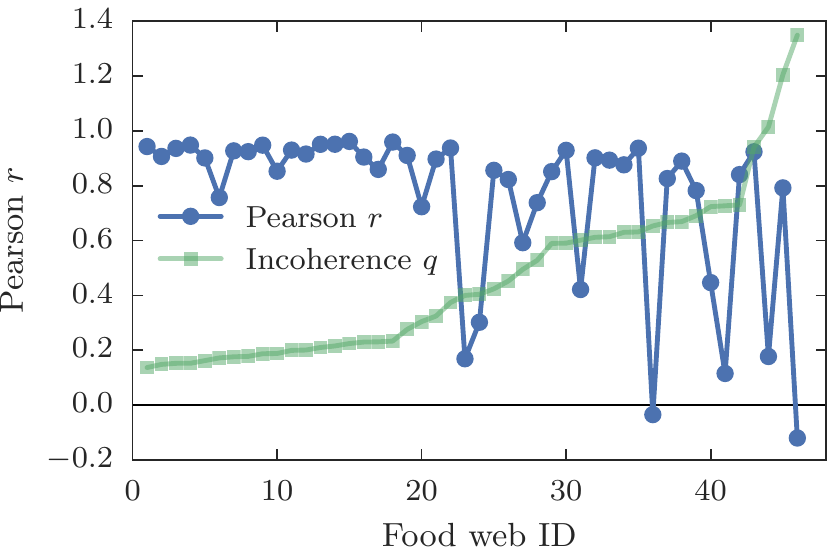}
  \caption{Pearson's correlation coefficient of the triad significance profiles (TSP). The coefficient is measured between the empirical TSP and the average TSP in the model ensemble over $1000$ simulated networks. Food webs are arranged by increasing incoherence parameter $q$.}
  \label{fig:r}
\end{figure}

\subsection{Comparison between empirical and model networks}\label{sec:model-empirical}
We have also investigated the similarities of triad significance profiles between the empirical food webs and model generated food webs. To this end we study the similarity of the TSPs between each empirical food web and an ensemble of model food webs fitted to the data of the empirical one. The results are summarized in Figure \ref{fig:r}. Averaging over an ensemble of $1000$ model generated food webs fitted to each empirical food web, we measured the Pearson correlation coefficient between the TSP of the empirical food web and the TSP of the ensemble average. The results show that for food webs in Family 1 (ID 1-22), the model is able to reproduce the empirical TSPs with high accuracy ($r\geq0.9$). For Family 2 (ID 23-46, excluding 30, 35, 40, 43), in most cases the same is true ($r\geq0.8$). The model fails to produce accurate TSPs for a number of food webs and sometimes even produces anti-correlated TSPs ($r<0$). These sporadic food webs correspond to the ones that generally do not qualify for membership in either family as shown
in the previous section.

\subsection{The role of omnivory and basal species}\label{sec:omnivory}
We now focus on the claim that the main difference between the two families of food webs is the relative under- and over-representation of triad S2, or the degree of omnivory in a food web. A prevalence of triad S2 indicates that the species in a food web often feed on different trophic levels, contributing to an increased incoherence parameter $q$ as discussed in Section \ref{sec:motifs}. A scarcity of triad S2, on the other hand, indicates that species only tend to feed on prey with similar trophic levels, which in turn signals a low incoherence parameter. This suggests a relationship between the $z$-score of triad S2 and network incoherence as measured by $q$.

Furthermore, model results in Section \ref{sec:model} suggest that a high proportion of basal species to all species, $B/N$, produces more coherent food webs (i.e. with a low incoherence parameter $q$). We take this as an additional predictive food web statistic for family membership.

\begin{figure}[htbp]
  \centering
  \centerline{\includegraphics[width=\linewidth]{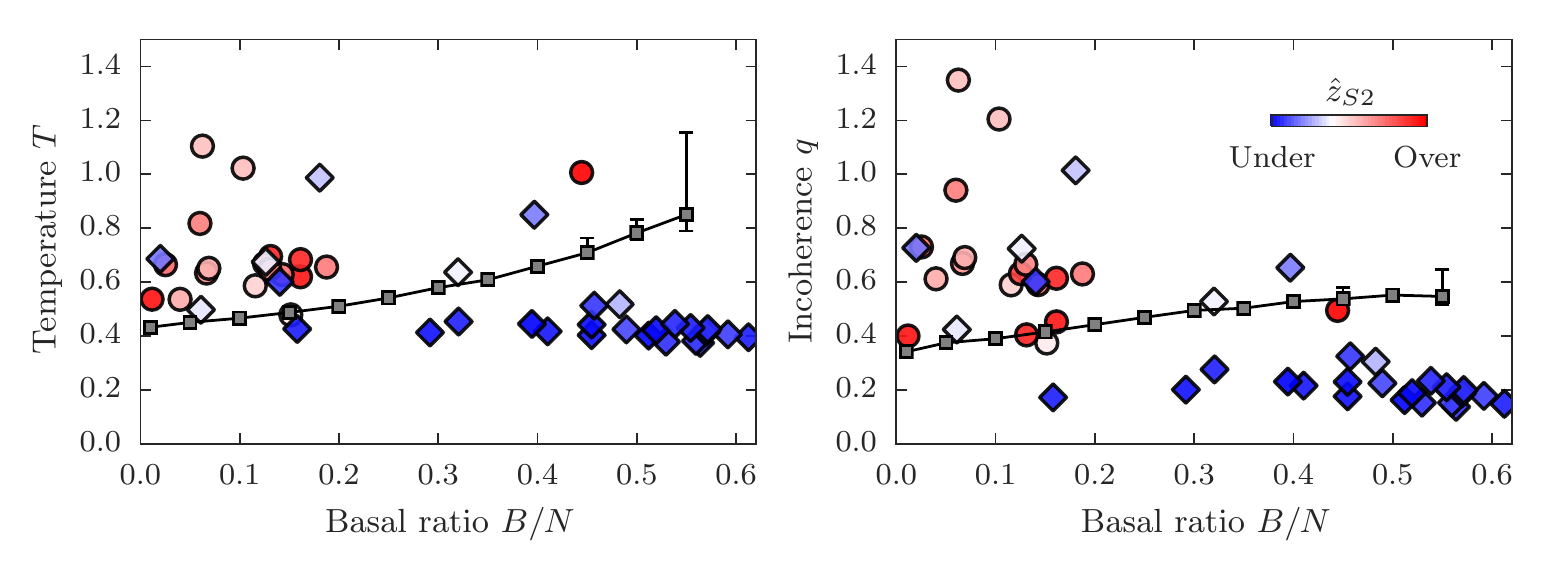}}
  \caption{Scatter plots of the temperature $T$ (left) and the incoherence parameter $q$ (right) versus the basal species ratio $B/N$ for all food webs. The gradient indicates the degree of over-representation (red circles) or under-representation (blue diamonds) of the feed-forward triad S2 as measured by the normalized z-score $\hat{z}_{\text{S2}}$. The line shows the transition from over-representation (above) to under-representation (below) as observed in the model with $N=100$, $\enmean{k}=L/(N-B)=10$ averaged over $100$ runs. Error bars are approximate 95\% confidence intervals.}
  \label{fig:comb_phase}
\end{figure}

Our findings are summarized in Figure \ref{fig:comb_phase}. This is a scatter plot of all $46$ food webs where we have plotted the fitted model temperature $T$ and the measured incoherence parameter $q$ against the ratio of basal species to all species $B/N$. We observe a clear anti-correlation ($R^2=-0.74$) that indicates a positive relationship between how coherent a network is (low $q$) and how many of its species are basal.

We have also coloured the markers of each food web to indicate the level of over- or under-representation of triad S2 as measured by the normalized $z$-score $\hat{z}_{\text{S2}}$. Red circles indicate an over-representation while blue diamonds indicate an under-representation of S2 in the respective food web. Remarkably, based on this measure, we uncover two clusters of food webs corresponding roughly to the two families based on TSP similarities. The first cluster is once again characterized by a high incoherence parameter $q$ as well as a low ratio of basal species to all species $B/N$. The second cluster is characterized by a low incoherence parameter and a high ratio of basal species to all species. The two clusters roughly correspond to the two families as characterized by the TSPs, so we can conclude that, indeed, the main difference between the two families is the relative role of triad S2 as already observed in the bulk behaviour of the TSPs in Figure \ref{fig:tsp}.

Finally, we study if our model exhibits a similar transition from a relatively S2-poor to an S2-rich state which would explain the relatively good agreement between empirical and model generated TSPs for the two families (Section \ref{sec:model-empirical}, Figure \ref{fig:r}). We find that for a given basal species ratio $B/N$ there exists a critical temperature $T_c$, and thus a critical incoherence parameter $q_c$, which signifies such a transition. For $T$ (and $q$) below these critical values, the model generates networks where S2 is under-represented, while for values above critical, the networks generated have either an even or an over-represented number of S2 triads. We include the transition line of the two regimes in Figure \ref{fig:comb_phase} for an ensemble of $100$ model networks with $N=100$ species and an average (non-basal) degree $\enmean{k}=L/(N-B)=10$. Networks with $q$ below the line show an under-representation of S2 triads, while networks with $q$ above the line show an over-representation as measured by $\hat{z}_{\text{S2}}$.

Remarkably, the model results are in very good agreement with the empirical data despite the fact that both the network size $N$ and the average degree $\enmean{k}$ vary between the empirical food webs. Almost all food webs with an under-represented number of S2 triads fall below the transition line of the model while those with an over-represented number reside above the line.

These findings are evidence that the distinction between the two families of food webs can be made based on the degree of omnivory present as measured by the prevalence of triad S2. Furthermore, this inherently local property is
intimately related to the global incoherence parameter $q$.  Interestingly, based on the strong anti-correlation between $q$ and $B/N$, the ratio of basal species to all species, both are equally good at determining which family a given food web belongs to. Finally, the realizability of networks belonging to either of the two families by our model suggests that the parameters $q$ and $B/N$ are both important in the mathematical modelling of food webs and may, in fact, be fundamental for understanding local preying patterns in food webs.

\section{Discussion}
Our investigation of trophic interaction patterns in food webs on two different scales reveals significant correlations between the degree of omnivory, hierarchical organization of trophic species and the density of basal species.

First, our analysis of local trophic interactions via triad significance profiles in empirical food webs reveals two distinct families of food webs characterized by a relatively low or high incoherence parameter respectively. While certain differences across families of food webs based on their TSPs have been observed before \cite{stouffer_evidence_2007}, these have not been examined satisfactorily in the context of structural differences in the food webs. Trophic coherence provides a global, network theoretic metric that enables us to classify and predict the relative prevalence of local trophic interactions.

Second, we show qualitatively that the the main difference between the two food web families is the degree of omnivory, as measured by the over- or under-representation of triad S2 (the ``feed-forward loop''). There is therefore a link between global order (trophic coherence) and local preying patterns.

Third, we test our prediction for the onset of omnivory using a new model for generating synthetic food webs with a given trophic coherence. We find that the model exhibits a transition from an under-representation of omnivory to an over-representation of omnivory as a function of trophic coherence. The model results fit the empirical food data very well, providing further evidence of the importance of capturing trophic coherence accurately in food web modelling.

Finally, we analyse the impact of the density of basal species prompted by our exploration of the model. We find that this parameter can be just as good a predictor of family membership as the trophic incoherence parameter.

To conclude, food web motifs, an inherently local property, can be understood in the context of trophic coherence -- a global food web property able to explain many features of food webs that have until now been at odds with established food web models \cite{johnson_trophic_2014}. We would like to emphasize that these findings are remarkably robust between food webs originating from vastly different habitats. Further work should focus on what mechanisms result in trophic coherence as a food web evolves, the specific role of basal species in determining the degree of omnivory in a given food web, and the qualitative differences in the handful of sporadic food webs that do not appear to fit well within this framework.

\section*{Acknowledgments}
JK was supported by the EPSRC under grant EP/IO1358X/1. SJ is grateful for support from Spanish MINECO Grant No. FIS2013-43201-P (FEDER funds).

\section*{Author contributions}
JK and SJ designed the research and wrote the paper. JK analysed the data and developed new tools.

\section*{Competing interests}
The authors declare no competing interests.

\section*{Data}
The datasets supporting this article have been uploaded as part of the supplementary material.








\section*{Food-web data}

\label{Section_Data}

We have compiled a dataset of 46 food webs available in the literature, pertaining to several ecosystem types. The methods used
by the researchers to establish the links between species vary from gut content analysis to inferences about the behaviour of similar
creatures. In Table \ref{table_foodwebs} we list the food webs used along with references to the relevant work. We also list, for each
case, the number of species $N$, of basal species $B$, of links $L$, the ecosystem type, the
incoherence parameter $q$, the value of the temperature parameter $T$ found to yield (on average) the empirical $q$ with our model, and the numerical ID
used to represent the food web in several figures in the main paper. Note that the values reported here exclude cannibalistic links which have a small
effect on the resulting values of $q$ and $T$.

\renewcommand{\tablename}{Table}
\setcounter{table}{0}
\setlength{\LTleft}{-20cm plus -1fill}
\setlength{\LTright}{\LTleft}
\begin{center}
 \begin{longtable}{@{}llccccccc}
\hline
  Food web & $N$ & $B$ & $L$ & $q$ & $T$ & Type & Reference &ID\\
\hline
Akatore Stream & 84 & 43 & 227 & 0.16 & 0.40 & River & \cite {streams5,streams6,streams7} & 5\\
Benguela Current & 29 & 2 & 196 & 0.69 & 0.65 & Marine &  \cite{benguela} & 39\\
Berwick Stream & 77 & 35 & 240 & 0.18 & 0.40 & River & \cite {streams5,streams6,streams7} & 7\\
Blackrock Stream & 86 & 49 & 375 & 0.19 & 0.42 & River & \cite {streams5,streams6,streams7} & 9\\
Bridge Broom Lake & 25 & 8 & 104 & 0.53 & 0.64 & Lake & \cite{bridge} & 28\\
Broad Stream & 94 & 53 & 564 & 0.14 & 0.37 & River & \cite {streams5,streams6,streams7} &1\\
Canton Creek & 102 & 54 & 696 & 0.15 & 0.38 & River & \cite{canton} & 4\\
Caribbean (2005) & 249 & 5 & 3302 & 0.73 & 0.69 & Marine & \cite{caribbean_2005} & 41\\
Caribbean Reef & 50 & 3 & 535 & 0.94 & 0.82 & Marine & \cite{reef} & 43\\
Carpinteria Salt Marsh Reserve & 126 & 50 & 541 & 0.65 & 0.85 & Marine & \cite{carpinteria} & 36\\
Caitlins Stream & 48 & 14 & 110 & 0.20 & 0.41 & River & \cite{streams5,streams6,streams7} & 12\\
Chesapeake Bay & 31 & 5 & 67 & 0.45 & 0.62 & Marine & \cite{chesapeake1,chesapeake2} & 26\\
Coachella Valley & 29 & 3 & 243 & 1.21 & 1.02 & Terrestrial & \cite{coachella} & 45\\
Coweeta (1) & 58 & 28 & 126 & 0.30 & 0.52 & River & \cite {streams5,streams6,streams7} & 20\\
Crystal Lake (Delta) & 19 & 3 & 30 & 0.17 & 0.43 & Lake & \cite{CrystalD} & 6\\
Cypress (Wet Season) & 64 & 12 & 439 & 0.63 & 0.66 & Terrestrial & \cite{south_florida98} & 34\\
Dempsters Stream (Autumn) & 83 & 46 & 414 & 0.21 & 0.43 & River & \cite {streams5,streams6,streams7} & 13\\
El Verde Rainforest & 155 & 28 & 1507 & 1.01 & 0.99 & Terrestrial & \cite{el_verde} & 44\\
Everglades Graminoid Marshes & 64 & 4 & 681 & 1.35 & 1.10 & Terrestrial & \cite{Everglades} & 46\\
Florida Bay & 121 & 14 & 1767 & 0.59 & 0.59 & Marine & \cite{south_florida98} & 29\\
German Stream & 84 & 48 & 352 & 0.20 & 0.43 & River & \cite {streams5,streams6,streams7} & 11\\
Grassland (U.K.) & 61 & 8 & 97 & 0.40 & 0.69 & River & \cite{grass} & 24\\
Healy Stream & 96 & 47 & 634 & 0.22 & 0.42 & River & \cite {streams5,streams6,streams7} & 15\\
Kyeburn Stream & 98 & 58 & 629 & 0.18 & 0.41 & River & \cite {streams5,streams6,streams7} & 8\\
LilKyeburn Stream & 78 & 42 & 375 & 0.23 & 0.44 & River & \cite {streams5,streams6,streams7} & 18\\
Little Rock Lake & 92 & 12 & 984 & 0.67 & 0.65 & Lake & \cite{little_rock} & 37\\
Lough Hyne & 349 & 49 & 5102 & 0.60 & 0.60 & Lake & \cite{lough_hyne_1,lough_hyne_2} & 31\\
Mangrove Estuary (Wet Season) & 90 & 6 & 1151 & 0.67 & 0.63 & Marine & \cite{south_florida98} & 38\\
Martins Stream & 105 & 48 & 343 & 0.32 & 0.51 & River & \cite {streams5,streams6,streams7} & 21\\
Maspalomas Pond & 18 & 8 & 24 & 0.49 & 1.01 & Lake & \cite{Maspalomas} & 27\\
Michigan Lake & 33 & 5 & 127 & 0.37 & 0.48 & Lake & \cite{Michigan} & 22\\
Narragansett Bay & 31 & 5 & 111 & 0.61 & 0.68 & Marine & \cite{Narragan} & 33\\
Narrowdale Stream & 71 & 28 & 154 & 0.23 & 0.44 & River & \cite {streams5,streams6,streams7} & 17\\
N.E. Shelf & 79 & 2 & 1378 & 0.73 & 0.66 & Marine & \cite{shelf} & 42\\
North Col Stream & 78 & 25 & 241 & 0.28 & 0.45 & River &  \cite {streams5,streams6,streams7} & 19\\
Powder Stream & 78 & 32 & 268 & 0.22 & 0.42 & River & \cite {streams5,streams6,streams7} &  14\\
Scotch Broom & 85 & 1 & 219 & 0.40 & 0.54 & Terrestrial & \cite{broom} & 23\\
Skipwith Pond & 25 & 1 & 189 & 0.61 & 0.54 & Lake & \cite{skipwith} & 32\\
St. Marks Estuary & 48 & 6 & 218 & 0.63 & 0.67 & Marine & \cite{st_marks} & 35\\
St. Martin Island & 42 & 6 & 205 & 0.59 & 0.63 & Terrestrial & \cite{st_martin} & 30\\
Stony Stream & 109 & 61 & 827 & 0.15 & 0.38 & River & \cite{stony} & 3\\
Sutton Stream (Autumn) & 80 & 49 & 335 & 0.15 & 0.40 & River & \cite {streams5,streams6,streams7} & 2\\
Troy Stream & 77 & 40 & 181 & 0.19 & 0.42 & River & \cite {streams5,streams6,streams7} & 10\\
Venlaw Stream & 66 & 30 & 187 & 0.23 & 0.44 & River & \cite {streams5,streams6,streams7} & 16\\
Weddell Sea & 483 & 61 & 15317 & 0.72 & 0.68 & Marine & \cite{weddell_sea} & 40\\
Ythan Estuary & 82 & 5 & 391 & 0.42 & 0.50 & Marine & \cite{Ythan96} & 25\\
\caption{
An alphabetical list of the 46 food webs used throughout the main paper. From left to right, the columns are for:
name, number of species $N$, number of basal species $B$, number of links $L$, ecosystem type, trophic
incoherence parameter $q$, value of the temperature parameter $T$ found to yield (on average) the empirical $q$ with our model, references to original work,
and the numerical ID.
}
\label{table_foodwebs}
 \end{longtable}
\end{center}



\bibliographystyle{prsb}
\bibliography{draft,si_draft}

\begin{thebibliography}{10}
\expandafter\ifx\csname urlstyle\endcsname\relax
  \providecommand{\doi}[1]{doi:\discretionary{}{}{}#1}\else
  \providecommand{\doi}{doi:\discretionary{}{}{}\begingroup
  \urlstyle{rm}\Url}\fi

\bibitem{Paine1966}
Paine RT, 1966 Food web complexity and species diversity.
\newblock \emph{The American Naturalist} \textbf{100}, 65--75.
\newblock (\doi{10.1086/282400})

\bibitem{Pimm1982}
Pimm SL, 1982 \emph{Food Webs}.
\newblock Springer Netherlands.
\newblock (\doi{10.1007/978-94-009-5925-5_1})

\bibitem{cohen2012community}
Cohen J, Briand F, Newman C, 1990 \emph{Community food webs: data and theory},
  vol.~20.
\newblock Berlin,Germany: Springer-Verlag.
\newblock (\doi{10.1007/978-3-642-83784-5})

\bibitem{dunne2004network}
Dunne JA, Williams RJ, Martinez ND, 2004 Network structure and robustness of
  marine food webs.
\newblock \emph{Marine Ecology Progress Series} \textbf{273}, 291--302.
\newblock (\doi{10.3354/meps273291})

\bibitem{drossel}
Drossel B, McKane AJ, 2005 Modelling food webs.
\newblock In \emph{Handbook of Graphs and Networks}, 218--247. Wiley-VCH Verlag
  GmbH \& Co. KGaA.
\newblock (\doi{10.1002/3527602755.ch10})

\bibitem{may1973stability}
May RM, 1973 \emph{Stability and complexity in model ecosystems}, vol.~6.
\newblock Princeton University Press

\bibitem{pimm_food_1991}
Pimm SL, Lawton JH, Cohen JE, 1991 Food web patterns and their consequences.
\newblock \emph{Nature} \textbf{350}, 669--674.
\newblock (\doi{10.1038/350669a0})

\bibitem{garlaschelli2003universal}
Garlaschelli D, Caldarelli G, Pietronero L, 2003 Universal scaling relations in
  food webs.
\newblock \emph{Nature} \textbf{423}, 165--168.
\newblock (\doi{10.1038/nature01604})

\bibitem{camacho_quantitative_2007}
Camacho J, Stouffer D, Amaral L, 2007 Quantitative analysis of the local
  structure of food webs.
\newblock \emph{Journal of theoretical biology} \textbf{246}, 260--268.
\newblock (\doi{10.1016/j.jtbi.2006.12.036})

\bibitem{paulau_motif_2015}
Paulau PV, Feenders C, Blasius B, 2015 Motif analysis in directed ordered
  networks and applications to food webs.
\newblock \emph{Scientific Reports} \textbf{5}, 11926.
\newblock (\doi{10.1038/srep11926})

\bibitem{borrelli_selection_2015}
Borrelli JJ, 2015 Selection against instability: stable subgraphs are most
  frequent in empirical food webs.
\newblock \emph{Oikos} \textbf{124}, 1583--1588.
\newblock (\doi{10.1111/oik.02176})

\bibitem{stouffer_evidence_2007}
Stouffer DB, Camacho J, Jiang W, Amaral LAN, 2007 Evidence for the existence of
  a robust pattern of prey selection in food webs.
\newblock \emph{Proceedings of the Royal Society of London B: Biological
  Sciences} \textbf{274}, 1931--1940.
\newblock (\doi{10.1098/rspb.2007.0571})

\bibitem{bascompte2005}
Bascompte J, Melián CJ, 2005 Simple trophic modules for complex food webs.
\newblock \emph{Ecology} \textbf{86}, 2868--2873.
\newblock (\doi{10.1890/05-0101})

\bibitem{itzkovitz_subgraphs_2003}
Itzkovitz S, Milo R, Kashtan N, Ziv G, Alon U, 2003 Subgraphs in random
  networks.
\newblock \emph{Physical Review E} \textbf{68}, 026127.
\newblock (\doi{10.1103/PhysRevE.68.026127})

\bibitem{milo_network_2002}
Milo R, Shen-Orr S, Itzkovitz S, Kashtan N, Chklovskii D, Alon U, 2002 Network
  motifs: simple building blocks of complex networks.
\newblock \emph{Science (New York, N.Y.)} \textbf{298}, 824--7.
\newblock (\doi{10.1126/science.298.5594.824})

\bibitem{milo_superfamilies_2004}
Milo R, Itzkovitz S, Kashtan N, Levitt R, Shen-Orr S, Ayzenshtat I, Sheffer M,
  Alon U, 2004 Superfamilies of evolved and designed networks.
\newblock \emph{Science (New York, N.Y.)} \textbf{303}, 1538--42.
\newblock (\doi{10.1126/science.1089167})

\bibitem{polis1991}
Polis GA, 1991 Complex trophic interactions in deserts: An empirical critique
  of food-web theory.
\newblock \emph{The American Naturalist} \textbf{138}, 123--155.
\newblock (\doi{10.1086/285208})

\bibitem{prill2005}
Prill RJ, Iglesias PA, Levchenko A, 2005 Dynamic properties of network motifs
  contribute to biological network organization.
\newblock \emph{PLoS Biol} \textbf{3}.
\newblock (\doi{10.1371/journal.pbio.0030343})

\bibitem{vazquez_topological_2004}
Vázquez A, Dobrin R, Sergi D, Eckmann JP, Oltvai ZN, Barabási AL, 2004 The
  topological relationship between the large-scale attributes and local
  interaction patterns of complex networks.
\newblock \emph{Proceedings of the National Academy of Sciences} \textbf{101},
  17940--17945.
\newblock (\doi{10.1073/pnas.0406024101})

\bibitem{dominguez-garcia_inherent_2014}
Domínguez-García V, Pigolotti S, Muñoz MA, 2014 Inherent directionality
  explains the lack of feedback loops in empirical networks.
\newblock \emph{Scientific Reports} \textbf{4}.
\newblock (\doi{10.1038/srep07497})

\bibitem{johnson_trophic_2014}
Johnson S, Domínguez-García V, Donetti L, Muñoz MA, 2014 Trophic coherence
  determines food-web stability.
\newblock \emph{Proceedings of the National Academy of Sciences} \textbf{111},
  17923--17928.
\newblock (\doi{10.1073/pnas.1409077111})

\bibitem{johnson_spectra_2015}
Johnson S, Jones NS, 2015 Spectra and cycle structure of trophically coherent
  graphs.
\newblock \emph{arXiv:1505.07332 [physics]} ArXiv: 1505.07332

\bibitem{dominguez-garcia_intervality_2016}
Domínguez-García V, Johnson S, Muñoz MA, 2016 Intervality and coherence in
  complex networks.
\newblock \emph{Chaos} \textbf{26}, 065308.
\newblock (\doi{10.1063/1.4953163})

\bibitem{klaise_neurons_2016}
Klaise J, Johnson S, 2016 From neurons to epidemics: How trophic coherence
  affects spreading processes.
\newblock \emph{Chaos} \textbf{26}, 065310.
\newblock (\doi{10.1063/1.4953160})

\bibitem{beber_artefacts_2012}
Beber ME, Fretter C, Jain S, Sonnenschein N, Müller-Hannemann M, Hütt MT,
  2012 Artefacts in statistical analyses of network motifs: general framework
  and application to metabolic networks.
\newblock \emph{Journal of The Royal Society Interface} \textbf{9}, 3426--3435.
\newblock (\doi{10.1098/rsif.2012.0490})

\bibitem{Newman2001}
Newman MEJ, Strogatz SH, Watts DJ, 2001 Random graphs with arbitrary degree
  distributions and their applications.
\newblock \emph{Phys. Rev. E} \textbf{64}, 026118.
\newblock (\doi{10.1103/PhysRevE.64.026118})

\bibitem{newman2010networks}
Newman M, 2010 \emph{Networks: An Introduction}.
\newblock New York, NY, USA: Oxford University Press, Inc.
\newblock (\doi{10.1093/acprof:oso/9780199206650.001.0001})

\bibitem{kashtan2002mfinder}
Kashtan N, Itzkovitz S, Milo R, Alon U, 2002 Mfinder tool guide.
\newblock \emph{Department of Molecular Cell Biology and Computer Science and
  Applied Math., Weizmann Inst. of Science, Rehovot Israel, technical report}

\bibitem{LEVINE1980195}
Levine S, 1980 Several measures of trophic structure applicable to complex food
  webs.
\newblock \emph{Journal of Theoretical Biology} \textbf{83}, 195 -- 207.
\newblock (\doi{10.1016/0022-5193(80)90288-X})

\bibitem{cohen1985}
Cohen J, Newman C, 1985 \emph{A stochastic theory of community food webs. I.
  Models and aggregated data.}, vol. 224, 421--448.
\newblock 1237 edn.
\newblock (\doi{10.1098/rspb.1985.0042})

\bibitem{williams2000simple}
Williams RJ, Martinez ND, 2000 Simple rules yield complex food webs.
\newblock \emph{Nature} \textbf{404}, 180--183.
\newblock (\doi{10.1038/35004572})

\bibitem{cattin2004phylogenetic}
Cattin MF, Bersier LF, Bana{\v{s}}ek-Richter C, Baltensperger R, Gabriel JP,
  2004 Phylogenetic constraints and adaptation explain food-web structure.
\newblock \emph{Nature} \textbf{427}, 835--839.
\newblock (\doi{10.1038/nature02327})

\bibitem{stouffer2005}
Stouffer DB, Camacho J, Guimerà R, Ng CA, Nunes~Amaral LA, 2005 Quantitative
  patterns in the structure of model and empirical food webs.
\newblock \emph{Ecology} \textbf{86}, 1301--1311.
\newblock (\doi{10.1890/04-0957})

\bibitem{stouffer2006}
Stouffer DB, Camacho J, Amaral LAN, 2006 A robust measure of food web
  intervality.
\newblock \emph{Proceedings of the National Academy of Sciences} \textbf{103},
  19015--19020.
\newblock (\doi{10.1073/pnas.0603844103})

\bibitem{allesina2008}
Allesina S, Alonso D, Pascual M, 2008 A general model for food web structure.
\newblock \emph{Science} \textbf{320}, 658--661.
\newblock (\doi{10.1126/science.1156269})

\bibitem{williams2008}
Williams RJ, Martinez ND, 2008 Success and its limits among structural models
  of complex food webs.
\newblock \emph{Journal of Animal Ecology} \textbf{77}, 512--519.
\newblock (\doi{10.1111/j.1365-2656.2008.01362.x})

\bibitem{streams5}
Thompson RM, Townsend CR, 2003 Impacts on stream food webs of native and exotic
  forest: An intercontinental comparison.
\newblock \emph{Ecology} \textbf{84}, 145--161.
\newblock (\doi{10.1890/0012-9658(2003)084[0145:iosfwo]2.0.co;2})

\bibitem{streams6}
Thompson RM, Townsend CR, 2005 Energy availability, spatial heterogeneity and
  ecosystem size predict food-web structure in stream.
\newblock \emph{Oikos} \textbf{108}, 137–148.
\newblock (\doi{10.1111/j.0030-1299.2005.11600.x})

\bibitem{streams7}
Townsend, Thompson, McIntosh, Kilroy, Edwards, Scarsbrook, 1998 Disturbance,
  resource supply, and food-web architecture in streams.
\newblock \emph{Ecology Letters} \textbf{1}, 200--209.
\newblock (\doi{10.1046/j.1461-0248.1998.00039.x})

\bibitem{benguela}
Yodzis P, 1998 Local trophodynamics and the interaction of marine mammals and
  fisheries in the benguela ecosystem.
\newblock \emph{Journal of Animal Ecology} \textbf{67}, 635--658.
\newblock (\doi{10.1046/j.1365-2656.1998.00224.x})

\bibitem{bridge}
Havens K, 1992 Scale and structure in natural food webs.
\newblock \emph{Science} \textbf{257}, 1107--1109.
\newblock (\doi{10.1126/science.257.5073.1107})

\bibitem{canton}
Townsend, Thompson, McIntosh, Kilroy, Edwards, Scarsbrook, 1998 Disturbance,
  resource supply, and food-web architecture in streams.
\newblock \emph{Ecology Letters} \textbf{1}, 200--209.
\newblock (\doi{10.1046/j.1461-0248.1998.00039.x})

\bibitem{caribbean_2005}
Bascompte J, Melián C, Sala E, 2005 Interaction strength combinations and the
  overfishing of a marine food web.
\newblock \emph{Proceedings of the National Academy of Sciences of the United
  States of America} \textbf{102}, 5443--5447.
\newblock (\doi{10.1073/pnas.0501562102})

\bibitem{reef}
Opitz S, 1996 Trophic interactions in {C}aribbean coral reefs.
\newblock \emph{ICLARM Tech. Rep.} \textbf{43}, 341

\bibitem{carpinteria}
Lafferty KD, Hechinger RF, Shaw JC, Whitney KL, Kuris AM, 2006 Food webs and
  parasites in a salt marsh ecosystem.
\newblock In SK~Collinge, C~Ray, eds., \emph{Disease ecology: Community
  structure and pathogen dynamics}, 119--134.
\newblock (\doi{10.1093/acprof:oso/9780198567080.003.0009})

\bibitem{chesapeake1}
Ulanowicz RE, Baird D, 1999 Nutrient controls on ecosystem dynamics: the
  {C}hesapeake mesohaline community.
\newblock \emph{Journal of Marine Systems} \textbf{19}, 159 -- 172.
\newblock (\doi{10.1016/S0924-7963(98)90017-3})

\bibitem{chesapeake2}
Abarca-Arenas LG, Ulanowicz RE, 2002 The effects of taxonomic aggregation on
  network analysis.
\newblock \emph{Ecological Modelling} \textbf{149}, 285--296.
\newblock (\doi{10.1016/S0304-3800(01)00474-4})

\bibitem{coachella}
Polis G, 1991 Complex trophic interactions in deserts: an empirical critique of
  food-web theory.
\newblock \emph{Am. Nat.} \textbf{138}, 123--125.
\newblock (\doi{10.1086/285208})

\bibitem{CrystalD}
Ulanowicz R, 1986 Growth and development: Ecosystems phenomenology. springer,
  new york. pp 69-79.
\newblock \emph{Network Analysis of Trophic Dynamics in South Florida
  Ecosystem, FY 97: The Florida Bay Ecosystem.}
  (\doi{10.1007/978-1-4612-4916-0})

\bibitem{south_florida98}
Ulanowicz R, Bondavalli C, Egnotovich M, 1998 Spatial and temporal variation in
  the structure of a freshwater food web.
\newblock \emph{Network Analysis of Trophic Dynamics in South Florida
  Ecosystem, FY 97: The Florida Bay Ecosystem.}

\bibitem{el_verde}
Waide RB, Reagan DP, 1996 \emph{The Food Web of a Tropical Rainforest}.
\newblock Chicago: University of Chicago Press

\bibitem{Everglades}
Ulanowicz R, Heymans J, Egnotovich M, 2000 Network analysis of trophic dynamics
  in south florida ecosystems.
\newblock \emph{Network Analysis of Trophic Dynamics in South Florida
  Ecosystems FY 99: The Graminoid Ecosystem.}

\bibitem{grass}
Martinez ND, Hawkins BA, Dawah HA, Feifarek BP, 1999 Effects of sampling effort
  on characterization of food-web structure.
\newblock \emph{Ecology} \textbf{80}, 1044–1055.
\newblock (\doi{10.1890/0012-9658(1999)080[1044:eoseoc]2.0.co;2})

\bibitem{little_rock}
Martinez ND, 1991 Artifacts or attributes? {E}ffects of resolution on the
  {L}ittle {R}ock {L}ake food web.
\newblock \emph{Ecol. Monogr.} \textbf{61}, 367--392.
\newblock (\doi{10.2307/2937047})

\bibitem{lough_hyne_1}
Riede J, Brose U, Ebenman B, Jacob U, Thompson R, Townsend C, Jonsson T, 2011
  Stepping in {E}lton's footprints: a general scaling model for body masses and
  trophic levels across ecosystems.
\newblock \emph{Ecology Letters} \textbf{14}, 169--178.
\newblock (\doi{10.1111/j.1461-0248.2010.01568.x})

\bibitem{lough_hyne_2}
Eklöf A, \emph{et~al.}, 2013 The dimensionality of ecological networks.
\newblock \emph{Ecology Letters} \textbf{16}, 577--583.
\newblock (\doi{10.1111/ele.12081})

\bibitem{Maspalomas}
Almunia J, Basterretxea G, Arısteguia J, Ulanowicz R, 1999 Benthic-pelagic
  switching in a coastal subtropical lagoon.
\newblock \emph{Estuarine, Coastal and Shelf Science} \textbf{49}, 363--384.
\newblock (\doi{10.1006/ecss.1999.0503})

\bibitem{Michigan}
Mason D, 2003 Quantifying the impact of exotic invertebrate invaders on food
  web structure and function in the great lakes: A network analysis approach.
\newblock \emph{Interim Progress Report to the Great Lakes Fisheries
  Commission- yr 1}

\bibitem{Narragan}
Monaco ME, Ulanowicz RE, 1997 Comparative ecosystem trophic structure of three
  u.s mid-atlantic estuaries.
\newblock \emph{Marine Ecology Progress Series} \textbf{161}, 239--254.
\newblock (\doi{10.3354/meps161239})

\bibitem{shelf}
Link J, 2002 Does food web theory work for marine ecosystems?
\newblock \emph{Mar. Ecol. Prog. Ser.} \textbf{230}, 1--9.
\newblock (\doi{10.3354/meps230001})

\bibitem{broom}
Memmott J, Martinez ND, Cohen JE, 2000 Predators, parasitoids and pathogens:
  species richness, trophic generality and body sizes in a natural food web.
\newblock \emph{J. Anim. Ecol.} \textbf{69}, 1--15.
\newblock (\doi{10.1046/j.1365-2656.2000.00367.x})

\bibitem{skipwith}
Warren PH, 1989 Spatial and temporal variation in the structure of a freshwater
  food web.
\newblock \emph{Oikos} \textbf{55}, 299--311.
\newblock (\doi{10.2307/3565588})

\bibitem{st_marks}
Christian RR, Luczkovich JJ, 1999 Organizing and understanding a winter's
  {S}eagrass foodweb network through effective trophic levels.
\newblock \emph{Ecol. Model.} \textbf{117}, 99--124.
\newblock (\doi{10.1016/S0304-3800(99)00022-8})

\bibitem{st_martin}
Goldwasser L, Roughgarden JA, 1993 Construction and analysis of a large
  {C}aribbean food web.
\newblock \emph{Ecology} \textbf{74}, 1216--1233.
\newblock (\doi{10.2307/1940492})

\bibitem{stony}
Townsend, Thompson, McIntosh, Kilroy, Edwards, Scarsbrook, 1998 Disturbance,
  resource supply, and food-web architecture in streams.
\newblock \emph{Ecology Letters} \textbf{1}, 200--209.
\newblock (\doi{10.1046/j.1461-0248.1998.00039.x})

\bibitem{weddell_sea}
Jacob U, \emph{et~al.}, 2011 The role of body size in complex food webs.
\newblock \emph{Advances in Ecological Research} \textbf{45}, 181--223.
\newblock (\doi{10.1016/b978-0-12-386475-8.00005-8})

\bibitem{Ythan96}
Huxham M, Beaney S, Raffaelli D, 1996 {Do parasites reduce the chances of
  triangulation in a real food web?}
\newblock \emph{Oikos} \textbf{76}, 284--300.
\newblock (\doi{10.2307/3546201})

\end{thebibliography}


\begin{thebibliography}{10}
\expandafter\ifx\csname urlstyle\endcsname\relax
  \providecommand{\doi}[1]{doi:\discretionary{}{}{}#1}\else
  \providecommand{\doi}{doi:\discretionary{}{}{}\begingroup
  \urlstyle{rm}\Url}\fi

\bibitem{streams5}
Thompson RM, Townsend CR, 2003 Impacts on stream food webs of native and exotic
  forest: An intercontinental comparison.
\newblock \emph{Ecology} \textbf{84}, 145--161.
\newblock (\doi{10.1890/0012-9658(2003)084[0145:iosfwo]2.0.co;2})

\bibitem{streams6}
Thompson RM, Townsend CR, 2005 Energy availability, spatial heterogeneity and
  ecosystem size predict food-web structure in stream.
\newblock \emph{Oikos} \textbf{108}, 137–148.
\newblock (\doi{10.1111/j.0030-1299.2005.11600.x})

\bibitem{streams7}
Townsend, Thompson, McIntosh, Kilroy, Edwards, Scarsbrook, 1998 Disturbance,
  resource supply, and food-web architecture in streams.
\newblock \emph{Ecology Letters} \textbf{1}, 200--209.
\newblock (\doi{10.1046/j.1461-0248.1998.00039.x})

\bibitem{benguela}
Yodzis P, 1998 Local trophodynamics and the interaction of marine mammals and
  fisheries in the benguela ecosystem.
\newblock \emph{Journal of Animal Ecology} \textbf{67}, 635--658.
\newblock (\doi{10.1046/j.1365-2656.1998.00224.x})

\bibitem{bridge}
Havens K, 1992 Scale and structure in natural food webs.
\newblock \emph{Science} \textbf{257}, 1107--1109.
\newblock (\doi{10.1126/science.257.5073.1107})

\bibitem{canton}
Townsend, Thompson, McIntosh, Kilroy, Edwards, Scarsbrook, 1998 Disturbance,
  resource supply, and food-web architecture in streams.
\newblock \emph{Ecology Letters} \textbf{1}, 200--209.
\newblock (\doi{10.1046/j.1461-0248.1998.00039.x})

\bibitem{caribbean_2005}
Bascompte J, Melián C, Sala E, 2005 Interaction strength combinations and the
  overfishing of a marine food web.
\newblock \emph{Proceedings of the National Academy of Sciences of the United
  States of America} \textbf{102}, 5443--5447.
\newblock (\doi{10.1073/pnas.0501562102})

\bibitem{reef}
Opitz S, 1996 Trophic interactions in {C}aribbean coral reefs.
\newblock \emph{ICLARM Tech. Rep.} \textbf{43}, 341

\bibitem{carpinteria}
Lafferty KD, Hechinger RF, Shaw JC, Whitney KL, Kuris AM, 2006 Food webs and
  parasites in a salt marsh ecosystem.
\newblock In SK~Collinge, C~Ray, eds., \emph{Disease ecology: Community
  structure and pathogen dynamics}, 119--134.
\newblock (\doi{10.1093/acprof:oso/9780198567080.003.0009})

\bibitem{chesapeake1}
Ulanowicz RE, Baird D, 1999 Nutrient controls on ecosystem dynamics: the
  {C}hesapeake mesohaline community.
\newblock \emph{Journal of Marine Systems} \textbf{19}, 159 -- 172.
\newblock (\doi{10.1016/S0924-7963(98)90017-3})

\bibitem{chesapeake2}
Abarca-Arenas LG, Ulanowicz RE, 2002 The effects of taxonomic aggregation on
  network analysis.
\newblock \emph{Ecological Modelling} \textbf{149}, 285--296.
\newblock (\doi{10.1016/S0304-3800(01)00474-4})

\bibitem{coachella}
Polis G, 1991 Complex trophic interactions in deserts: an empirical critique of
  food-web theory.
\newblock \emph{Am. Nat.} \textbf{138}, 123--125.
\newblock (\doi{10.1086/285208})

\bibitem{CrystalD}
Ulanowicz R, 1986 Growth and development: Ecosystems phenomenology. springer,
  new york. pp 69-79.
\newblock \emph{Network Analysis of Trophic Dynamics in South Florida
  Ecosystem, FY 97: The Florida Bay Ecosystem.}
  (\doi{10.1007/978-1-4612-4916-0})

\bibitem{south_florida98}
Ulanowicz R, Bondavalli C, Egnotovich M, 1998 Spatial and temporal variation in
  the structure of a freshwater food web.
\newblock \emph{Network Analysis of Trophic Dynamics in South Florida
  Ecosystem, FY 97: The Florida Bay Ecosystem.}

\bibitem{el_verde}
Waide RB, Reagan DP, 1996 \emph{The Food Web of a Tropical Rainforest}.
\newblock Chicago: University of Chicago Press

\bibitem{Everglades}
Ulanowicz R, Heymans J, Egnotovich M, 2000 Network analysis of trophic dynamics
  in south florida ecosystems.
\newblock \emph{Network Analysis of Trophic Dynamics in South Florida
  Ecosystems FY 99: The Graminoid Ecosystem.}

\bibitem{grass}
Martinez ND, Hawkins BA, Dawah HA, Feifarek BP, 1999 Effects of sampling effort
  on characterization of food-web structure.
\newblock \emph{Ecology} \textbf{80}, 1044–1055.
\newblock (\doi{10.1890/0012-9658(1999)080[1044:eoseoc]2.0.co;2})

\bibitem{little_rock}
Martinez ND, 1991 Artifacts or attributes? {E}ffects of resolution on the
  {L}ittle {R}ock {L}ake food web.
\newblock \emph{Ecol. Monogr.} \textbf{61}, 367--392.
\newblock (\doi{10.2307/2937047})

\bibitem{lough_hyne_1}
Riede J, Brose U, Ebenman B, Jacob U, Thompson R, Townsend C, Jonsson T, 2011
  Stepping in {E}lton's footprints: a general scaling model for body masses and
  trophic levels across ecosystems.
\newblock \emph{Ecology Letters} \textbf{14}, 169--178.
\newblock (\doi{10.1111/j.1461-0248.2010.01568.x})

\bibitem{lough_hyne_2}
Eklöf A, \emph{et~al.}, 2013 The dimensionality of ecological networks.
\newblock \emph{Ecology Letters} \textbf{16}, 577--583.
\newblock (\doi{10.1111/ele.12081})

\bibitem{Maspalomas}
Almunia J, Basterretxea G, Arısteguia J, Ulanowicz R, 1999 Benthic-pelagic
  switching in a coastal subtropical lagoon.
\newblock \emph{Estuarine, Coastal and Shelf Science} \textbf{49}, 363--384.
\newblock (\doi{10.1006/ecss.1999.0503})

\bibitem{Michigan}
Mason D, 2003 Quantifying the impact of exotic invertebrate invaders on food
  web structure and function in the great lakes: A network analysis approach.
\newblock \emph{Interim Progress Report to the Great Lakes Fisheries
  Commission- yr 1}

\bibitem{Narragan}
Monaco ME, Ulanowicz RE, 1997 Comparative ecosystem trophic structure of three
  u.s mid-atlantic estuaries.
\newblock \emph{Marine Ecology Progress Series} \textbf{161}, 239--254.
\newblock (\doi{10.3354/meps161239})

\bibitem{shelf}
Link J, 2002 Does food web theory work for marine ecosystems?
\newblock \emph{Mar. Ecol. Prog. Ser.} \textbf{230}, 1--9.
\newblock (\doi{10.3354/meps230001})

\bibitem{broom}
Memmott J, Martinez ND, Cohen JE, 2000 Predators, parasitoids and pathogens:
  species richness, trophic generality and body sizes in a natural food web.
\newblock \emph{J. Anim. Ecol.} \textbf{69}, 1--15.
\newblock (\doi{10.1046/j.1365-2656.2000.00367.x})

\bibitem{skipwith}
Warren PH, 1989 Spatial and temporal variation in the structure of a freshwater
  food web.
\newblock \emph{Oikos} \textbf{55}, 299--311.
\newblock (\doi{10.2307/3565588})

\bibitem{st_marks}
Christian RR, Luczkovich JJ, 1999 Organizing and understanding a winter's
  {S}eagrass foodweb network through effective trophic levels.
\newblock \emph{Ecol. Model.} \textbf{117}, 99--124.
\newblock (\doi{10.1016/S0304-3800(99)00022-8})

\bibitem{st_martin}
Goldwasser L, Roughgarden JA, 1993 Construction and analysis of a large
  {C}aribbean food web.
\newblock \emph{Ecology} \textbf{74}, 1216--1233.
\newblock (\doi{10.2307/1940492})

\bibitem{stony}
Townsend, Thompson, McIntosh, Kilroy, Edwards, Scarsbrook, 1998 Disturbance,
  resource supply, and food-web architecture in streams.
\newblock \emph{Ecology Letters} \textbf{1}, 200--209.
\newblock (\doi{10.1046/j.1461-0248.1998.00039.x})

\bibitem{weddell_sea}
Jacob U, \emph{et~al.}, 2011 The role of body size in complex food webs.
\newblock \emph{Advances in Ecological Research} \textbf{45}, 181--223.
\newblock (\doi{10.1016/b978-0-12-386475-8.00005-8})

\bibitem{Ythan96}
Huxham M, Beaney S, Raffaelli D, 1996 {Do parasites reduce the chances of
  triangulation in a real food web?}
\newblock \emph{Oikos} \textbf{76}, 284--300.
\newblock (\doi{10.2307/3546201})

\end{thebibliography}

\end{document}